\documentclass[lettersize,journal]{IEEEtran}
\usepackage{amsmath,amsfonts}
\usepackage{algorithmic}
\usepackage{algorithm}
\usepackage{array}
\usepackage[caption=false,font=normalsize,labelfont=sf,textfont=sf]{subfig}
\usepackage{textcomp}
\usepackage{stfloats}
\usepackage{url}
\usepackage{verbatim}
\usepackage{graphicx}
\usepackage{cite}
\usepackage{physics}
\begin{document}

\title{Quantum Transfer Learning for Real-World, Small, and High-Dimensional Remotely Sensed Datasets}

\author{Soronzonbold Otgonbaatar, Gottfried Schwarz,  Mihai Datcu, \IEEEmembership{Fellow, IEEE}, and Dieter Kranzlmüller
\thanks{Soronzonbold Otgonbaatar is with Remote Sensing Technology Institute, German Aerospace Center
DLR, Weßling, 82234, Germany and Ludwig-Maximilians-Universität München, Gottfried Schwarz and Mihai Datcu are with the Remote Sensing Technology Institute, German Aerospace Center
DLR, Weßling, 82234, Germany, and Dieter Kranzlmüller is with the Ludwig-Maximilians-Universität München.}
\thanks{Corresponding author: S. Otgonbaatar (Soronzonbold.Otgonbaatar@dlr.de)}
}

\markboth{IEEE Journal of Selected Topics in Applied Earth Observations and Remote Sensing}%
{Shell \MakeLowercase{\textit{et al.}}: A Sample Article Using IEEEtran.cls for IEEE Journals}


\maketitle

\begin{abstract}
Quantum Machine Learning (QML) models promise to have some computational (or quantum) advantage for classifying supervised datasets (e.g., satellite images) over some conventional Deep Learning (DL) techniques due to their expressive power via their local effective dimension. There are, however, two main challenges regardless of the promised quantum advantage: 1) Currently available quantum bits (qubits) are very small in number, while real-world datasets are characterized by hundreds of high-dimensional elements (\emph{i.e.} features). Additionally, there is not a single unified approach for embedding real-world high-dimensional datasets in a limited number of qubits. 2) Some real-world datasets are too small for training intricate QML networks. Hence, to tackle these two challenges for benchmarking and validating QML networks on real-world, small, and high-dimensional datasets in one-go, we employ quantum transfer learning comprising a classical VGG16 layer and a multi-qubit QML layer. We use real-amplitude and strongly-entangling N-layer QML networks with and without data re-uploading layers as a multi-qubit QML layer, and evaluate their expressive power quantified by using their local effective dimension; the lower the local effective dimension of a QML network, the better its performance on unseen data. As datasets, we utilize Eurosat and synthetic datasets (\emph{i.e.} easy-to-classify datasets), and an UC Merced Land Use dataset (\emph{i.e.} a hard-to-classify dataset). Our numerical results show that the strongly-entangling N-layer QML network has a lower local effective dimension than the real-amplitude QML network and outperforms it on the hard-to-classify datasets. In addition, quantum transfer learning helps tackle the two challenges mentioned above for benchmarking and validating QML networks on real-world, small, and high-dimensional datasets. 
\end{abstract}


\section{Introduction}
\IEEEPARstart{U}{niversal} quantum computers are composed of a collection of quantum bits (qubits) and parameterized quantum gates being arranged according to some given topologies, while quantum learning algorithms are algorithms manipulating qubits by using parameterized quantum gates. Based on the learnable parameters of parameterized quantum gates, Quantum Machine Learning (QML) as described by \cite{biamonte} and \cite{qpca}, in general, contains three different sub-directions \cite{dunjko}: 
\begin{figure}[!t]   
 \includegraphics[width=\columnwidth]{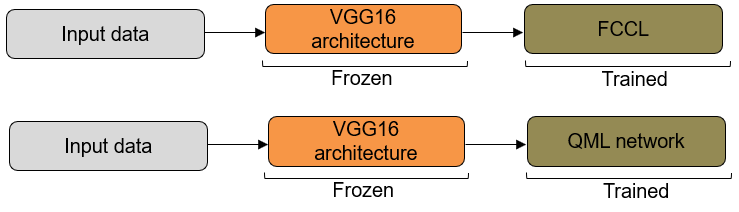}
  \caption{[Top] classical transfer learning: Input data, a DL VGG16 architecture, a Fully-Connected Classical Layer (FCCL), [Bottom] quantum transfer learning: Input data, a DL VGG16 architecture, a QML network.}\label{fig: qtl}
\end{figure}
\begin{itemize}
    \item \textbf{Quantum-Inspired} ML: develop novel artificial intelligence (AI/DL) techniques by using concepts from quantum information processing \cite{qinspired} and \cite{Felser_qinspired}.
    \item \textbf{Quantum} ML  (sometimes): apply classical DL techniques to quantum data (quantum chemistry) \cite{Zheng2021}.
    \item \textbf{Quantum-Applied} ML: develop quantum learning circuits for supervised real-world data on quantum computers \cite{WILLSCH2020}, \cite{Tuysuz2021}, and \cite{sozogate}.
\end{itemize}

In this work, we focus on Quantum-Applied ML, and we use QML interchangeably with Quantum-Applied ML. A QML network encodes input data in qubits, and learns the parameters of parameterized quantum gates. Moreover, it promises quantum advantage for some computational problems over conventional learning methods due to either its expressive power measured by the local effective dimension \cite{Abbas2021} and \cite{abbaslocal}, or its computational time \cite{qsvm}.

However, the qubits of currently available universal quantum computers are noisy and small in number. Hence, these types of universal quantum computers are called Noisy Intermediate-Scale Quantum (NISQ) computers \cite{preskill}. Due to the limited number of qubits, there are two main challenges:
\begin{enumerate}
    \item {Embedding challenge}: there is not a single unified approach for embedding real-world, high-dimensional data points in a small number of qubits.
    \item {Small dataset challenge}: QML networks do not capture informative patterns in small datasets in contrast to big datasets, and this challenge even exists with conventional DL methods.
\end{enumerate}
Here, we name a dataset as a low-dimensional dataset if and only if we can represent its data points by at most five elements using a classical dimensionality-reduction technique, and as a high-dimensional dataset otherwise. 

To overcome the \emph{embedding} challenge, some studies already proposed an embedding strategy for a toy dataset \cite{schuld}, as well as for real-world, big, and low-dimensional datasets \cite{sozogate}. In the work of \cite{sozogate}, its authors investigated a binary-label \emph{classic-quantum} classifier for embedding and classifying a two-label, low-dimensional Eurosat dataset \cite{eurosat}, in which they classified this specific binary dataset by measuring directly the output of the quantum layer. On the other hand, the article \cite{copy1} focused on a multi-label \emph{classic-quantum-classic} classifier in which the authors classified the same Eurosat dataset by measuring the last \emph{classical layer} but not a \emph{quantum layer} for a multi-class case. Furthermore, the authors of \cite{sozotgrs} already also introduced a single- and multi-qubit quantum classifier for embedding a selected practical dataset in a parameterized quantum circuit. 


To overcome the \emph{small dataset} challenge, the authors of \cite{schuld_qtf} proposed a novel method named \emph{quantum transfer learning}, the quantum version of classical transfer learning. Classical transfer learning is a procedure for training fully-connected classical layers residing on the top level of conventional DL architectures (for small datasets, when the parameters of frozen conventional DL architectures are initialized a priori on big similar datasets). 
In contrast, quantum transfer learning trains a QML network placed on the top level of frozen conventional DL architectures instead of fully-connected classical layers (see Fig. \ref{fig: qtl}). In particular, for quantum transfer learning, one replaces the fully-connected top layers of frozen conventional DL architectures by a multi-qubit QML network. One also profits from the advantage of quantum transfer learning since it simultaneously helps tackle the two challenges mentioned above. 

For practically important datasets, the authors of \cite{sozogate} and \cite{copy1} \emph{implicitly} used a quantum transfer learning method, but they employed a real-world, \emph{big}, and \emph{low-dimensional} dataset (\emph{i.e.} an Eurosat dataset). In this work, we \emph{explicitly} propose and employ a quantum transfer learning method for real-world, \emph{small}, and \emph{high-dimensional} datasets. Our proposed quantum transfer learning method consists of a multi-qubit QML network (with or without \emph{data re-uploading}), and a very deep convolutional network (in our case, a VGG16 architecture), playing the role of a feature extractor from datasets as shown in Fig. \ref{fig: qtl} [Bottom] \cite{datareuploading, seth}, and \cite{vgg16}. In particular, we employ real-amplitude and strongly-entangling N-layer QML networks with and without \emph{data re-uploading} layers \cite{schuldnlayer} as a multi-qubit QML network, and quantify their expressive power by using their \emph{local effective dimension} (the lower, the better) which gives us a portion of the active parameters in the trained QML network \cite{Abbas2021, Haug_2021,abbaslocal, classicallocal} and \cite{yuxuan_2020} and currently {available quantum resources}. The expressive power is referred to as the capacity of any learning model, and its capability to capture intricate patterns in any dataset. Moreover, the lower the local effective dimension of a given QML network, the better its performance on still unseen data points. 

Our practical datasets are synthetic, Eurosat, and UC Merced Land Use images  \cite{cheng0}. First, we compute the local effective dimension of our QML networks, that is, real-amplitude and strongly-entangling N-layer QML networks without \emph{data re-uploading}  layers, on two-class low-dimensional synthetic and Eurosat datasets since we can compress and represent these low-dimensional datasets by $3$ and $4$ elements using a classical DL network as proposed by \cite{sozogate}, respectively; here, we  generated and used two-class synthetic data including $100$ data points, where each data point is a two-dimensional vector drawn from circular data with the error according to a normal distribution $\mathcal{N}(\mu=0, \sigma=1)$. For low-dimensional Eurosat data, we utilized its \emph{annual crop} and \emph{residential} area classes, and this two-class set consists of $4,000$ data points, each of which is characterized by $64\times 64\times 3$ low-dimensional elements. To validate the relationship between local effective dimension and performance (\emph{i.e.} the classification accuracy) of a given QML network, we trained, subsequently, our QML networks via quantum transfer learning on the small, high-dimensional, and hard-to-classify three-class images, \emph{i.e.} \emph{dense residential}, \emph{medium residential}, and \emph{sparse residential} area classes of the high-dimensional UC Merced Land Use dataset. Additionally, a \emph{dense residential}, \emph{medium residential}, and \emph{sparse residential} area classification problem meets our two challenges mentioned, because this three-class images consist of only $288$ data points, and each image is characterized by $256\times 256\times 3$ high-dimensional elements \cite{cheng0} and \cite{cheng1}. As a quantum simulator, we used a PennyLane Python library for training our QML networks \cite{pennylane}.

Our experimental results demonstrate that the strongly-entangling N-layer QML network without \emph{data re-uploading}  layers has a lower local effective dimension than the real-amplitude N-layer QML network without \emph{data re-uploading} layers. It also has a higher test accuracy on real-world datasets than the real-amplitude QML network. Furthermore, quantum transfer learning helps tackle the two above-mentioned challenges for benchmarking and validating different QML networks on real-world, small, high-dimensional, and hard-to-classify datasets. 

This work is structured as follows: in Sections \ref{sec: our model} and \ref{sec: ourqml}, we provide some background for quantum transfer learning and multi-qubit QML networks, respectively. In Section \ref{ch: powerqnn}, we present the expressive power of QML networks via their local effective dimension. Subsequently, in Section \ref{sec: our data}, we introduce practical datasets being used in this paper. In Section \ref{sec: benchmark}, we present our experiments and some of our findings. Finally, we draw a few conclusions in Section \ref{sec: conclude}.

\section{quantum transfer learning}\label{sec: our model}
Quantum transfer learning is referred to as training a QML network with and without \emph{data re-uploading} on real-world small datasets when the weights of the VGG16 architecture are initialized on the \emph{ImageNet} dataset. Moreover, we froze the weights of the VGG16 such that none of its weights were updated during the training of a QML network. 
The QML network we propose in this study is a multi-qubit N-layer quantum classifier (with and without \emph{data re-uploading}); these classifiers are extremely simple as well as very powerful learning networks for non-linear datasets \cite{datareuploading} and \cite{seth}.

\subsection{Single-Qubit QML Network with and without Data Re-Uploading}
We characterize a single-qubit QML network by a tensorial feature map and universal parameterized quantum gates:

\begin{equation}\label{eq: gate}
    U(\phi_1, \phi_2, \phi_3)V_i(\theta_{i,1}, \gamma_{i,2}, \phi_{i,3}).
\end{equation}
where $\phi_1, \phi_2,$ and $\phi_3$ embed input data points in qubits, and $\theta_{i,1}, \gamma_{i,2},$ and $\phi_{i,3}$ are learning weights at the $i$th layer. Namely, we encode an input data point having three features in qubits using a tensorial feature map, $U(\phi_1, \phi_2, \phi_3)=U(\phi_1)\otimes U(\phi_2)\otimes U(\phi_3)$, and subsequently, we train a parameterized quantum gate $V_i(\theta_{i,1}, \gamma_{i,2}, \phi_{i,3})$ on the embedded data points.
For simplicity, we express $U=U(\phi_1, \phi_2, \phi_3)$ and $V_i=V_i(\theta_{i,1}, \gamma_{i,2}, \phi_{i,3})=V_i(\theta_{i,1})V_i(\gamma_{i,2})V_i(\phi_{i,3})$ since they are unitary quantum gates such that $V_i=V_i^\dagger, i=1, \dots, N$ where N is the number of layers or the depth of a given QML network \cite{sozotgrs}. In matrix form, we express $U$ and the $V_i$'s by:

\begin{equation}\label{eq: u}
    \begin{split}
    U(\phi_j) &=e^{i\phi_j}\begin{pmatrix}
        \cos(\phi_j) & -\sin(\phi_j)\\
        \sin(\phi_j) & \cos(\phi_j)
    \end{pmatrix}, \quad j =1,2,3,
    \end{split}
\end{equation}
and

\begin{equation}\label{eq: v1}
    \begin{split}
    V_i(\theta_{i,1})&=\begin{pmatrix}
        e^{i\theta_{i,1}} & 0\\
        0 & e^{-i\theta_{i,1}}
    \end{pmatrix},
    \quad
    V_i(\gamma_{i,2})=\begin{pmatrix}
        e^{i\gamma_{i,2}} & 0\\
        0 & e^{-i\gamma_{i,2}}
    \end{pmatrix},
    \end{split}
\end{equation}
and

\begin{equation}
    V_i(\phi_{i,3}) =\begin{pmatrix}
        \cos(\phi_{i,3}) & -\sin(\phi_{i,3})\\
        \sin(\phi_{i,3}) & \cos(\phi_{i,3})
    \end{pmatrix}, \quad i=1, \dots, N.
\end{equation}

\begin{figure}[!t]   
 \includegraphics[width=\columnwidth]{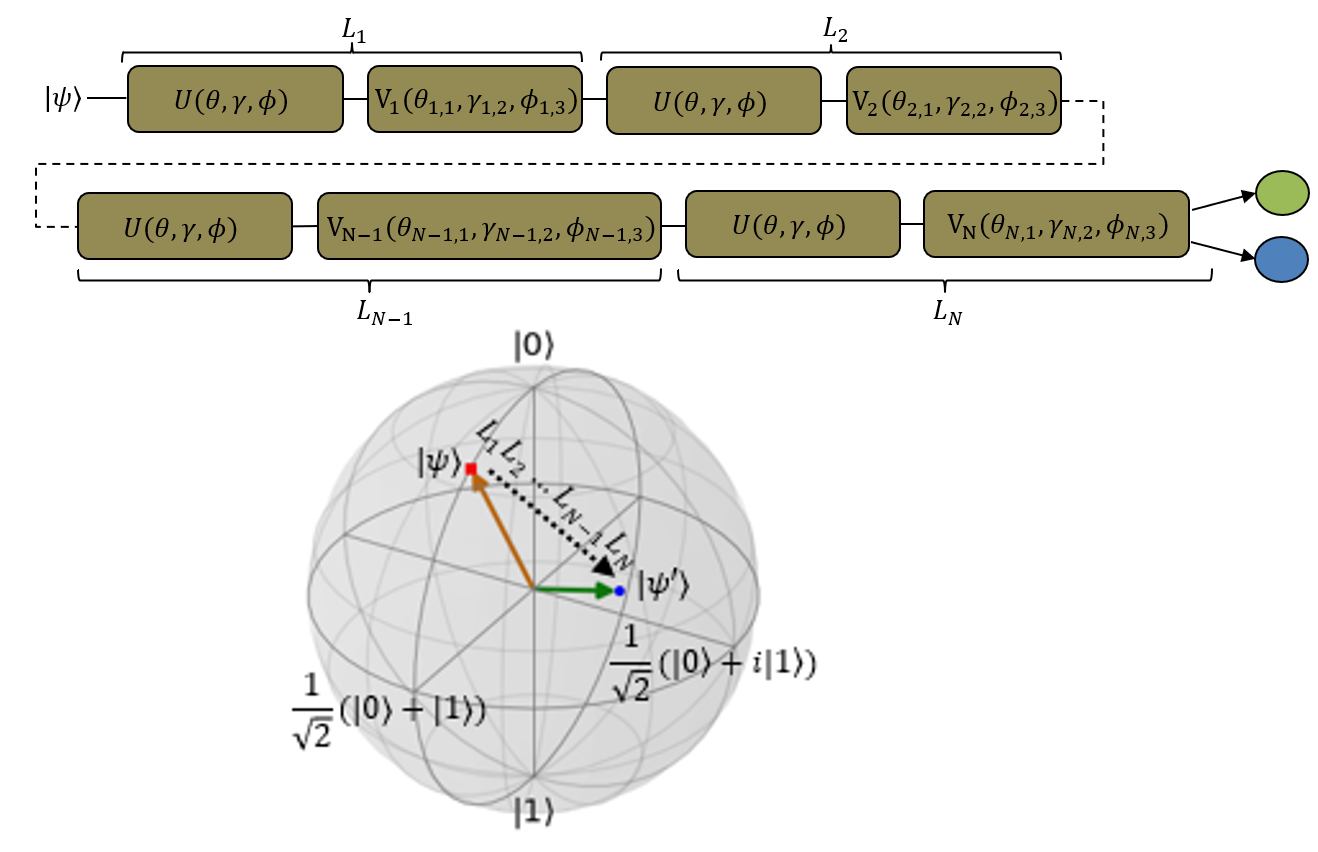}
  \caption{[Top] A single-qubit N-layer QML network with \emph{data re-uploading}, [Bottom] A QML network operation on a single-qubit represented on a Bloch sphere, in which $\ket{\psi}$ is an input single-qubit, and $\ket{\psi^\prime}$ is the single-qubit after $L_1, L_2, \dots, L_{N-1}, L_N$ quantum layers.}\label{fig: ournlayers}
\end{figure}

For a QML network with \emph{data re-uploading}, the quantum circuit $UV_i$ expressed by Eq. \eqref{eq: gate} is repeated $N$ times: 

\begin{equation}
\prod_{i=1}^N U V_i=\underbrace{U V_1}_{L_1} \underbrace{UV_2}_{L_2}\dots \underbrace{U V_{N-1}}_{L_{N-1}}\underbrace{U V_N}_{L_N},
\end{equation}
where an input data point is re-uploaded N times in the quantum gate $U$, and the parameterized quantum gate $V_N$ is also repeated N times as shown in Fig. \ref{fig: ournlayers}. The N\textsuperscript{th} layer is denoted as $L_N$. 

\begin{figure}[!t]   
 \includegraphics[width=\columnwidth]{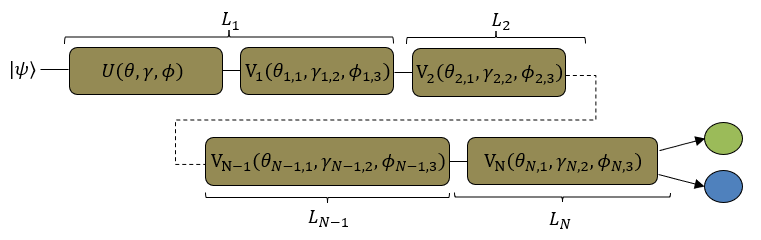}
  \caption{A single-qubit N-layer QML network without \emph{data re-uploading}}\label{fig: ournlayers_nore}
\end{figure}

Alternatively, for a QML network without \emph{data re-uploading}, the quantum circuit $UV_i$ expressed in Eq. \eqref{eq: gate} is repeated N times: 

\begin{equation}
U\prod_{i=1}^N V_i=\underbrace{U V_1}_{L_1} \underbrace{V_2}_{L_2}\dots \underbrace{V_{N-1}}_{L_{N-1}}\underbrace{V_N}_{L_N},
\end{equation}
where an input data point is uploaded in $U$ only once at layer $L_1$ as shown in Fig. \ref{fig: ournlayers_nore}.

For single-qubit QML networks, a qubit is represented by:

\begin{equation}\label{eq: ket}
\ket{0}=
\begin{pmatrix}
    1\\
    0
\end{pmatrix}, \quad 
\ket{1}=
\begin{pmatrix}
    0\\
    1
\end{pmatrix},
\end{equation}
and as a bra vector $\bra{0}=\ket{0}^\dagger, \bra{1}=\ket{1}^\dagger$, where $^\dagger$ represents both transpose and conjugate. In general, qubits can exist in a superposition:

\begin{equation}
    \ket{\psi}=c_1\ket{0}+c_2\ket{1} \quad \textit{such that} \quad |c_1|^2+|c_2|^2=1,    
\end{equation}
where $c_1$ and $c_2$ are complex numbers as shown in Fig. \ref{fig: ournlayers} [Bottom].

Let us consider a simple example for a single-qubit \emph{data re-uploading} QML network: we assume $\ket{\psi}=\ket{1}$ and a QML network with one layer $L_1$, $(0, 0, \phi_3)=(0,0,\pi/2)$ and $(\theta_{1,1}, \gamma_{1,2}, \phi_{1,3})=(0,0,0)$. A qubit after layer $L_1$ becomes:

\begin{equation}
    \ket{\psi^\prime}=V_1(0,0,0) U(\pi/2) \ket{1}.
\end{equation}

If we measure a single-qubit by using a projective measurement $\hat{z}=\dyad{0}-\dyad{1}$, (\emph{i.e.} the outer product of $\ket{0}$ and $\ket{1}$), we obtain an expected value of measuring the state $\ket{\psi^\prime}$ in the basis $\hat{z}$:

\begin{equation}
\begin{split}
     \expval{\hat{z}}_{\Vec{\theta}}&=\expval{\hat{z}}{\psi^\prime}= e^{-i\pi}e^{i\pi}\bra{0}\hat{z}\ket{0}=1.
\end{split}
\end{equation}
where $\Vec{\theta}=(\theta_{i,1}, \gamma_{i,2}, \phi_{i, 3})^T \in \Theta$ and $T$ denotes the transpose operation of a vector. In the end, this expected value is connected to two classical neurons shown in Fig. \ref{fig: ournlayers}, and these neurons denoted as $l=1,2$ are defined by: 

\begin{equation}\label{eq: 2n}
    \tilde{y}_l=A(a_l+w_l \expval{\hat{z}}_{\Vec{\theta}}),
\end{equation}
where $A(\cdot)$ is a non-linear activation function, that is, a sigmoid function, $\tilde{y}_l$ is their predicted output, $a_l$ is their bias, and $w_l$ is their edge parameter \cite{sozotgrs}.

In general, we express the expected value of a single-qubit N-layer QML network with and without \emph{data re-uploading} into the  basis $\hat{z}=\dyad{0}-\dyad{1}$ by:

\begin{equation}
    \expval{\hat{z}}_{\Vec{\theta}}=\expval{L_1^\dagger L_{2}^\dagger\dots L_{N-1}^\dagger L_N^\dagger \hat{z}  L_N L_{N-1}\dots L_2 L_1}{\psi}
\end{equation}
yielding a continuous value from $-1$ to $+1$. We then connect two classical neurons to the expected value $\expval{\hat{z}}_{\Vec{\theta}}$ as shown in Eq. \eqref{eq: 2n}.

\section{our multi-qubit qml networks with and without data re-uploading}\label{sec: ourqml}
A multi-qubit QML network is a parameterized quantum circuit (PQC) with some input qubits and parameterized quantum gates exploiting Eq. \eqref{eq: u} and  Eq. \eqref{eq: v1}. We encoded our data in its input qubits by utilizing a tensorial feature map expressed by Eq. \eqref{eq: u} as proposed by \cite{schuld,datareuploading}, and \cite{seth}, and we trained its parameterized quantum gates following Eq. \eqref{eq: v1}. Moreover, we name a multi-qubit QML network with N layers as a multi-qubit N-layer QML network similar to the single-qubit N-layer QML network mentioned in the previous section.

\begin{figure}[!t]   
 \includegraphics[width=\columnwidth]{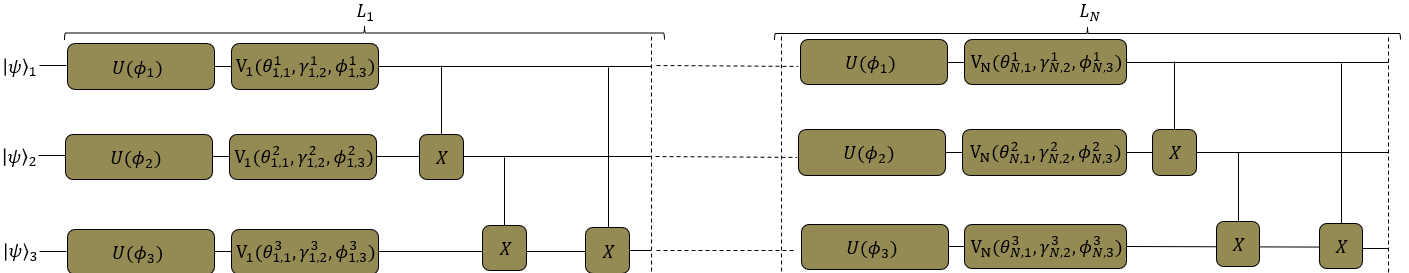}
  \caption{A multi-qubit N-layer QML network with \emph{data re-uploading}: $L_1, \dots, L_N$ quantum layers.}\label{fig: ournlayers_cnot}
\end{figure}

\begin{figure}[!t]   
 \includegraphics[width=\columnwidth]{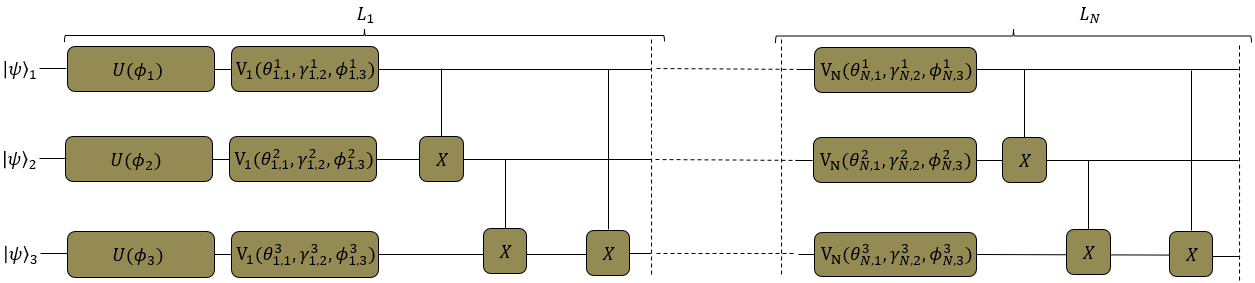}
  \caption{A multi-qubit N-layer QML network without \emph{data re-uploading}: $L_1, \dots, L_N$ quantum layers.}\label{fig: ournlayers_cnot_noU}
\end{figure}

In this work, we use three-qubit real-amplitude and strongly-entangling N-layer QML networks with and without \emph{data re-uploading} layers. We express the parameterized quantum gates of these QML networks with \emph{data re-uploading} layers by:

\begin{equation}
\prod_{i=1}^N\underbrace{U(\phi_1, \phi_2, \phi_3)V_i(\theta^q_{i,1}, \gamma^q_{i,2}, \phi^q_{i,3}) W_i}_{L_i}, \quad q=1,2,3,
\end{equation}
where $q$ represents the qubit number $\ket{\psi}_q$, and each $W_i$ denotes controlled-X quantum gates (see Fig. \ref{fig: ournlayers_cnot}). A controlled-X quantum gate is a two-qubit quantum gate acting on a target qubit if and only if a control qubit is in the state $\ket{1}$.
In contrast, these QML networks without \emph{data re-uploading} layers are characterized by: 

\begin{equation}
U(\phi_1, \phi_2, \phi_3)\prod_{i=1}^N\underbrace{V_i(\theta^q_{i,1}, \gamma^q_{i,2}, \phi^q_{i,3}) W_i}_{L_i},
\end{equation}
where the classical data is uploaded once at the layer $L_1$ as illustrated in Fig. \ref{fig: ournlayers_cnot_noU}.

\subsection{Real-Amplitude N-Layer QML Networks with and without Data Re-Uploading Layers}
Our real-amplitude N-layer QML networks with and without \emph{data re-uploading} layers are composed of a quantum gate $U$ encoding a data point in qubits, parameterized quantum gates $V_i$, and non-parameterized quantum gates $W_i$, respectively: 

\begin{equation}
\prod_{i=1}^N\underbrace{U(\phi_1, \phi_2, \phi_3)V_i(0, 0, \phi^q_{i,3}) W_i}_{L_i}
\end{equation}
and
\begin{equation}
U(\phi_1, \phi_2, \phi_3)\prod_{i=1}^N\underbrace{V_i(0, 0, \phi^q_{i,3}) W_i}_{L_i},
\end{equation}
where $W_i$ represents all-to-all entanglement \cite{copy1}.

\subsection{Strongly-Entangling N-Layer QML Networks with and without Data Re-Uploading Layers}
Our strongly-entangling N-layer QML networks with and without \emph{data re-uploading} layers are defined by: 

\begin{equation}
\prod_{i=1}^N\underbrace{U(\phi_1, \phi_2, \phi_3)V_i(\theta^q_{i,1}, \gamma^q_{i,2}, \phi^q_{i,3}) W_i}_{L_i}
\end{equation}
and
\begin{equation}
U(\phi_1, \phi_2, \phi_3)\prod_{i=1}^N\underbrace{V_i(\theta^q_{i,1}, \gamma^q_{i,2}, \phi^q_{i,3}) W_i}_{L_i},
\end{equation}
where $W_i$ represents strong entanglement \cite{schuldnlayer}.

\section{the power of quantum machine learning networks}\label{ch: powerqnn}


Deep neural networks are powerful learning models for identifying intricate patterns in big datasets. Their power is quantified by using the so-called local effective dimension which yields a portion of active parameters in the trained network \cite{abbaslocal} and \cite{classicallocal}. Quantum neural networks, \emph{i.e.} QML networks, are novel learning models based on PQCs exploiting quantum superposition, entanglement, and the interference of qubits.
The authors of \cite{Abbas2021} and \cite{abbaslocal} demonstrated that some QML networks have a lower local effective dimension $-$ a lower local effective dimension and being more powerful $-$ for analyzing some data-driven tasks than their classical counterparts. In particular, the local effective dimension gives us the active parameters $\Vec{\theta}^*$ in the trained model, and the lower effective dimension of a learning model (classical or quantum), the better it generalizes on unseen data points.

According to \cite{Abbas2021, abbaslocal}, and \cite{Meyer_2021}, the local effective dimension of a QML network $\expval{\hat{z}}_{\Vec{\theta}}$ around the active parameters $\Vec{\theta}^*\in\Theta\subset\mathbb{R}^d$ with $n$ training data points, where $\Theta$ weight space, is defined by:

\begin{equation}
    d_{n,\lambda}(\expval{\hat{z}}_{\Vec{\theta}}) = \frac{2\log \left(\dfrac{1}{V_\epsilon}\int_{\mathcal{B}_\epsilon(\Vec{\theta}^*)} \sqrt{det(I_d+k_{n,\lambda}\Bar{F}(\Vec{\theta})}d\Vec{\theta}\right)}{\log k_{n,\lambda}},
\end{equation}
where $\mathcal{B}_\epsilon(\Vec{\theta}^*) := \{\Vec{\theta}\in \Theta: \norm{\Vec{\theta}-\Vec{\theta}^*}\leq\epsilon\}$ is an $\epsilon$-ball with a volume $V_\epsilon$, $\epsilon>1/\sqrt{n}$, $k_{n,\lambda} = \dfrac{\lambda n}{2\pi\log n},  \lambda\in \left(\dfrac{2\pi\log n}{n}, 1\right]$,  $I_d$ is a unit diagonal matrix, and $\Bar{F}(\Vec{\theta})\in \mathbb{R}^{d\times d}$ is the normalized Fisher information matrix of $F(\Vec{\theta})$ \cite{FIM} with an element: 

\begin{equation}
    \Bar{F}_{ij}(\Vec{\theta})=\dfrac{d\cdot V_\epsilon}{\int_{\mathcal{B}_\epsilon(\Vec{\theta}^*)}\Tr(F(\Vec{\theta}))d\Vec{\theta}}F_{ij}(\Vec{\theta}).
\end{equation}

\section{our datasets}\label{sec: our data}

We generated and used a synthetic dataset using 
\begin{equation}
    x_m = r_m \begin{pmatrix}
        \cos\phi_m \\
        \sin\phi_m
    \end{pmatrix}+\begin{pmatrix}
        \epsilon_m^x\\
        \epsilon_m^y
    \end{pmatrix}
\end{equation}
where $r_m=1$ if $y_m=1$, and $r_m=0.15$ if $y_m=0$, $\epsilon_m$'s are a normal distribution $\mathcal{N}(\mu=0, \sigma=1)$, and $\phi_n \in (0,2\pi]$ linearly spaced \cite{sozomdpi}. This synthetic dataset is a two-class dataset composed of $m=100$ data points.

\begin{figure}[t!]   
 \includegraphics[width=\columnwidth]{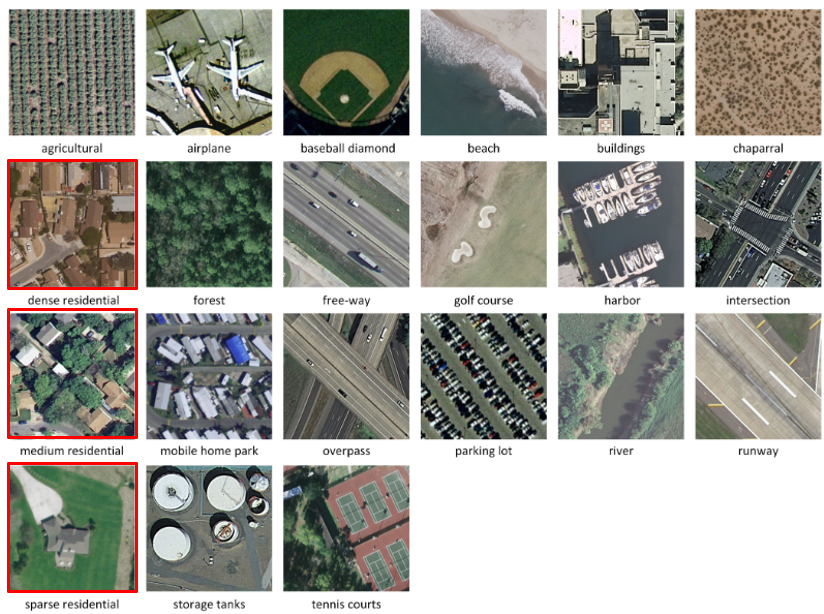}
  \caption{Some examples of the high-dimensional UC Merced Land Use dataset taken from \cite{cheng0}. Here, we contoured its hard-to-classify three-class examples in red.}\label{fig: ucmerced}
\end{figure}
\begin{figure}[t!]   
 \includegraphics[width=\columnwidth, height= 1.6cm]{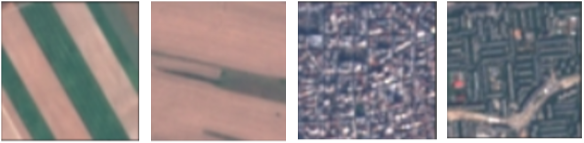}
  \caption{\emph{Annual crop} and \emph{residential} area classes of the low-dimensional Eurosat dataset.}\label{fig: eurosat}
\end{figure}

A Eurosat dataset is a Sentinel-2 image dataset with $27,000$ labelled and georeferenced images. Additionally, this dataset is a patch-based dataset with $64\times 64$ pixels comprising $10$ classes, where each image is characterized by $13$ spectral bands ranging from $443$ nm to $2190$ nm, and having a spatial resolution of $60$ m/pixel (see Fig. \ref{fig: eurosat}). 
We used selected two-class image scenes, namely the \emph{annual crop} and \emph{residential} area classes, consisting of $4,000$ images each of which is characterized by $64\times 64 \times 3$ low-dimensional elements.

UC Merced Land Use data contain image scenes of 21 classes with three RGB spectral bands. Its smallest class, \emph{tennis court}, includes $42$ data points, while its largest classes (e.g., \emph{agricultural}), include $100$ data points. In total, this dataset comprises $2100$ image scenes \cite{cheng0, cheng1}. By visual inspection and experiment, we learned that the hard-to-classify three-class examples of this dataset are the \emph{dense residential}, \emph{medium residential}, and \emph{sparse residential} area classes shown in Fig. \ref{fig: ucmerced}. For our tests, we employed this hard-to-classify three-class examples comprising $288$ image scenes, where each image is a $256\times 256\times 3$ high-dimensional elements.

\begin{figure}[t!]   
\centering
 \includegraphics[width=\columnwidth, height=6cm]{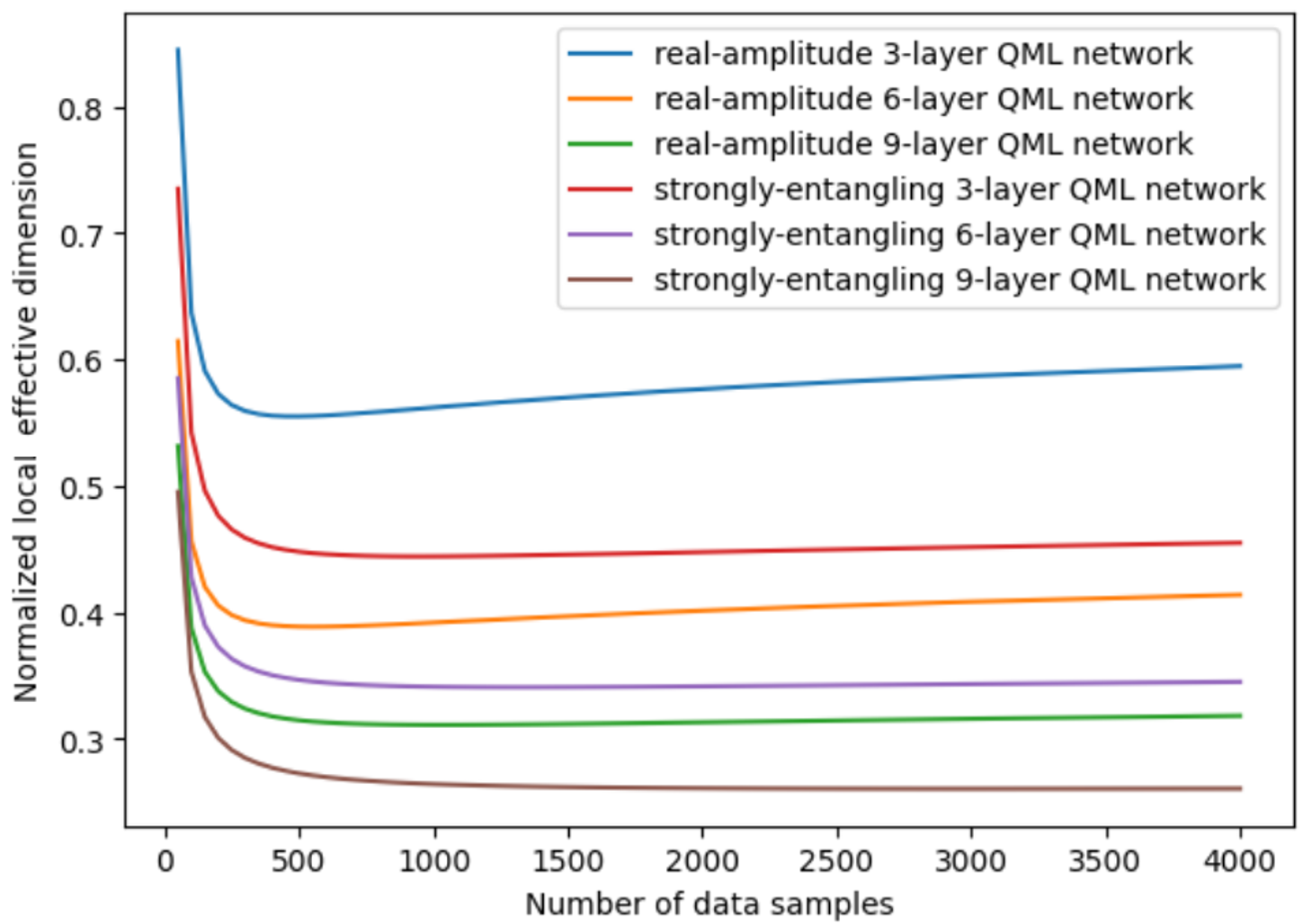}
  \caption{Normalized local effective dimension of real-amplitude and strongly-entangling N-layer QML networks without \emph{data re-uploading} with respect to the number of data samples.}\label{fig: leff}
\end{figure}

\section{our experiments}\label{sec: benchmark}

In this section, we define the local effective dimension and the classification performance of our QML networks, that is, real-amplitude and strongly-entangling N-layer QML networks, while utilizing synthetic and Eurosat datasets. In particular, we derive the relationship between the local effective dimension and the classification accuracy of our QML networks on synthetic and real-world datasets. 
In order to validate the relationship between the local effective dimension and the classification power of our QML networks, 
we train them via quantum transfer learning on the real-world, small, and high-dimensional three-class dataset (the hard-to-classify dataset), \emph{i.e.} \emph{dense residential}, \emph{medium residential}, and \emph{sparse residential} area classes, of the UC Merced Land Use dataset, because this three-class dataset meets the two challenges mentioned before: 
\begin{enumerate}
    \item {Embedding challenge}: the three-class images are characterized by $256\times 256\times 3$ high-dimensional elements, while currently available quantum computers already have around $50$ qubits.
    \item {Small dataset challenge}: the three-class dataset consists of only $288$ image scenes, and we split it into a training dataset comprising $201$ image scenes, and a test dataset consisting of $87$ image scenes.
\end{enumerate}

In particular, we benchmarked and validated our QML networks placed on the top level of the frozen VGG16 network on the hard-to-classify three-class dataset, because these datasets play a very important role in QML \cite{Huang2021}. 

\subsection{Local Effective Dimension and Performance of our QML Networks without Data Re-Uploading}
In our numerical study, we employed N-layer QML networks with only three and four qubits in order to keep the quantum resources as low as possible when $N=3, 6,$ or $9$. For the synthetic dataset, we computed the local effective dimension of real-amplitude and strongly-entangling N-layer QML networks without \emph{data re-uploading} while increasing the size of the dataset. We plotted the local effective dimension of our QML networks with respect to the dataset size shown in Fig. \ref{fig: leff} and present their classification accuracy in Table \ref{Table: benchmark_synth}. This result leads to the conclusion that the local effective dimension of QML networks seems to correlate with their classification accuracy. Furthermore, we discovered that strongly-entangling N-layer QML networks have a lower effective dimension and, at the same time, a higher accuracy for generating two-class labels than their counterpart real-amplitude QML networks.

\begin{table}[t!]
\caption{classification accuracy of real-amplitude and strongly-entangling n-layer qml networks without data re-uploading on the two-class synthetic dataset}
\label{Table: benchmark_synth}
\centering
\begin{tabular}{|c|c|c|c|c|c|c|}
 \hline
\multicolumn{1}{|c|}{}&\multicolumn{3}{|c|}{real-amplitude}& \multicolumn{3}{|c|}{strongly-entangling}\\ 
 \hline
 \multicolumn{1}{|c|}{\{Class\}/N-layer}&\multicolumn{1}{|c|}{$3$}&\multicolumn{1}{|c|}{$6$}&\multicolumn{1}{|c|}{$9$} &\multicolumn{1}{|c|}{$3$}&\multicolumn{1}{|c|}{$6$}&\multicolumn{1}{|c|}{$9$}\\ 
 \hline
$\{1,2\}$ & $0.35$ &  $0.45$  & $0.40$ & $ 0.90$ & $0.95$ & $1.00$\\
 \hline 
\end{tabular}
\end{table}
\begin{table}[!t]
\caption{Classification accuracy of $N$-depth real-amplitude and strongly-entangling networks {without a data re-uploading layer} on the annual crop and residential area classes of the Eurosat dataset; here, $\{1,2\}$ represents the annual crop and residential area classes.}
\label{Table: benchmark_eurosat}
\centering
\begin{tabular}{|c|c|c|c|c|c|c|}
 \hline
\multicolumn{1}{|c|}{}&\multicolumn{3}{|c|}{real-amplitude}& \multicolumn{3}{|c|}{strongly-entangling}\\ 
 \hline
 \multicolumn{1}{|c|}{\{Class\}/$N$-depth}&\multicolumn{1}{|c|}{$3$}&\multicolumn{1}{|c|}{$6$}&\multicolumn{1}{|c|}{$9$} &\multicolumn{1}{|c|}{$3$}&\multicolumn{1}{|c|}{$6$}&\multicolumn{1}{|c|}{$9$}\\ 
 \hline
$\{1,2\}$ & $0.70$ &  $0.73$  & $0.75$ & $ 0.74$ & $0.72$ & $0.70$\\
 \hline 
\end{tabular}
\end{table}
\begin{table}[!t]
\caption{Classification accuracy of $N$-depth real-amplitude and strongly-entangling networks {without a data re-uploading layer} on the dense residential, medium residential, and sparse residential area classes of the UC Merced Land Use dataset; here, $\{1,2,3\}$ represents the dense residential, medium residential, and sparse residential area classes.}
\label{Table: benchmark1_uc}
\centering
\begin{tabular}{|c|c|c|c|c|c|c|}
 \hline
\multicolumn{1}{|c|}{}&\multicolumn{3}{|c|}{real-amplitude}& \multicolumn{3}{|c|}{strongly-entangling}\\ 
 \hline
 \multicolumn{1}{|c|}{\{Class\}/$N$-depth}&\multicolumn{1}{|c|}{$3$}&\multicolumn{1}{|c|}{$6$}&\multicolumn{1}{|c|}{$9$} &\multicolumn{1}{|c|}{$3$}&\multicolumn{1}{|c|}{$6$}&\multicolumn{1}{|c|}{$9$}\\ 
 \hline
$\{1,2,3\}$ & $0.64$ &  $0.67$  & $0.72$ & $ 0.81$ & $0.83$ & $0.83$\\
 \hline 
\end{tabular}
\end{table}
\begin{table}[!t]
\caption{Classification accuracy of $N$-depth real-amplitude and strongly-entangling networks {with a data re-uploading layer} on the dense residential, medium residential, and sparse residential area classes of the UC Merced Land Use dataset; here, $\{1,2,3\}$ represents the dense residential, medium residential, and sparse residential area classes.}
\label{Table: benchmark2_uc}
\centering
\begin{tabular}{|c|c|c|c|c|c|c|}
 \hline
\multicolumn{1}{|c|}{}&\multicolumn{3}{|c|}{real-amplitude}& \multicolumn{3}{|c|}{strongly-entangling}\\ 
 \hline
 \multicolumn{1}{|c|}{\{Class\}/$N$-depth}&\multicolumn{1}{|c|}{$3$}&\multicolumn{1}{|c|}{$6$}&\multicolumn{1}{|c|}{$9$} &\multicolumn{1}{|c|}{$3$}&\multicolumn{1}{|c|}{$6$}&\multicolumn{1}{|c|}{$9$}\\ 
 \hline
$\{1,2, 3\}$ & $0.41$ &  $0.52$  & $0.48$ & $ 0.74$ & $0.66$ & $0.74$\\
 \hline 
\end{tabular}
\end{table}

To validate this conclusion for a real-world, big, and low-dimensional dataset comprising the \emph{annual crop} and \emph{residential} area classes of the Eurosat dataset, we first applied a so-called two-level encoding which maps each image scene characterized by $64\times64\times 3$ low-dimensional elements to $2\times2$ informative features using the VGG16 architecture, and encoded these informative features in four qubits by employing a tensorial feature map \cite{sozogate}. Then we calculated numerically the local effective dimension of the real-amplitude and strongly-entangling N-layer QML networks with a $3200$-element training set and an $800$-element testing set randomly sampled from the \emph{annual crop} and \emph{residential} area classes. We found that the strongly-entangling N-layer QML network generates two-class labels similar (and even better in some instances) to the ones generated by the real-amplitude N-layer QML network (see Table \ref{Table: benchmark_eurosat}), because the \emph{annual crop} and \emph{residential} area classes have less overlap which is proven by generating two-class labels and by visual inspection. In our case, this two-class classification problem is an easy-to-classify two-class labelling problem, though the strongly-entangling N-layer QML network has a lower local effective dimension and is more powerful than the real-amplitude N-layer QML network.

\begin{algorithm}[!t]
\caption{Quantum transfer learning for the hard-to-classify three-class images on the PennyLane simulator}\label{Table: qtf_classifier}
\begin{algorithmic}[1]
\STATE \textsc{input:} The hard-to-classify three-class examples 
\STATE \textsc{output:} The hard-to-classify three-class labels
\STATE \textsc{quantum transfer learning}: a sequential model having the frozen VGG16 layer for extracting informative features from remotely-sensed datasets, and a three-qubit QML layer for training on the output of the frozen VGG16 layer (see Fig. \ref{fig: qtl} [Bottom])
\STATE \textsc{training parameters:} epochs=20, batch=64, and the Adam optimizer having the learning rate of $10^{-4}$.
\STATE \textsc{stop algorithm}
\end{algorithmic}
\end{algorithm}

\subsection{Quantum Transfer Learning for Real-World, Small, High-Dimensional, and Hard-to-Classify Three-Class Images}

We validated the relationship between the local effective dimension and the classification accuracy of our QML networks with and without \emph{data re-uploading} on the real-world, small, and high-dimensional three-class images (\emph{i.e.} hard-to-classify images) of the UC Merced Land Use dataset, that is, \emph{dense residential}, \emph{medium residential}, and \emph{sparse residential} area classes. To validate this relationship illustrated in Fig. \ref{fig: leff}, we placed our QML networks with three input qubits on the top layer of the VGG16 network, since we applied our hard-to-classify three-class images split into a $201$-element training set and an $87$-element testing set, where each data point is characterized by $256\times256\times3$ high-dimensional elements compared with the big, low-dimensional Eurosat dataset. We summarized our results in Tables \ref{Table: benchmark1_uc} and \ref{Table: benchmark2_uc}. Here, we could prove that strongly-entangling N-layer networks without \emph{data re-uploading} layers outperform real-amplitude N-layer networks without \emph{data re-uploading} layers in most instances, except for the $N=9$ case due to their lower local effective dimension (see Table \ref{Table: benchmark1_uc} and Fig. \ref{fig: leff}). 
The poor performance of the strongly-entangling 9-layer QML network without \emph{data re-uploading} layers is caused by their vanishing gradient \cite{McClean2018}, that is beyond the main scope of this article but an important future research direction. Moreover, real-amplitude and strongly-entangling N-layer networks without \emph{data re-uploading} layers outperform (on practical high-dimensional datasets) ones with \emph{data re-uploading} layers on the same dataset (see Tables \ref{Table: benchmark1_uc} and \ref{Table: benchmark2_uc}).

\section{conclusion}\label{sec: conclude} 

We employed and benchmarked real-amplitude and strongly-entangling N-layer QML networks with and without \emph{data re-uploading} layers. As practical datasets, we used a two-class synthetic dataset, easy-to-classify two-class images of the Eurosat dataset, and  hard-to-classify three-class images of the UC Merced Land Use dataset. The hard-to-classify three-class dataset consists of $288$ image scenes, where each image scene has $256\times 256\times 3$ high-dimensional elements, while the easy-to-classify two-class images are composed of $4,000$ image scenes each of which is characterized by $64\times 64\times 3$ low-dimensional elements. In particular, our hard-to-classify three-class images meet the above-mentioned two challenges for training multi-qubit QML networks: the embedding and the small dataset challenge. 

We analyzed the expressive power of real-amplitude and strongly-entangling N-layer QML networks without \emph{data re-uploading} layers via their so-called local effective dimension, while utilizing the synthetic and Eurosat datasets. Our numerical experiments proved the statement that the lower the local effective dimension of a multi-qubit QML network, the better its classification accuracy on unseen data points. More importantly, we discovered that our strongly-entangling N-layer QML networks have a lower local effective dimension and a higher test accuracy than real-amplitude QML networks (see Fig. \ref{fig: leff} and Table \ref{Table: benchmark_synth}). Here, the local effective dimension of QML networks seems to correlate with their classification performance. We note, however, that for easy-to-classify datasets (in our case, two-class labels of the low-dimensional Eurosat dataset), multi-qubit QML networks perform equally well on unseen data points even though one of them is more powerful than other ones (see Table \ref{Table: benchmark_eurosat}). Thus, the hard-to-classify datasets are very important datasets for benchmarking and validating QML networks \cite{Huang2021}. 

To validate the relationship between the local effective dimension and the classification performance of real-amplitude and strongly-entangling N-layer QML networks, we trained them via quantum transfer learning on the real-world, small, and hard-to-classify three-class images of the high-dimensional UC Merced Land Use dataset. Our experimental results demonstrate that strongly-entangling N-layer QML networks perform better than real-amplitude N-layer QML networks in most instances (see Table \ref{Table: benchmark1_uc}). Furthermore, real-amplitude and strongly-entangling N-layer QML networks without \emph{data re-uploading} outperform ones with \emph{data re-uploading}. More importantly, quantum transfer learning even helps tackle the two main challenges in one-go encountered for benchmarking and validating multi-qubit QML networks on real-world, small, and high-dimensional datasets of practical importance. 

We must note, however, that our message is that we did not attempt to demonstrate computational advantage of QML models over their conventional counterparts, which is already demonstrated by the authors of \cite{sozogate}, \cite{Abbas2021}, \cite{abbaslocal}, and \cite{Haug_2021} but design and select a powerful model among existing QML models for real-world problems of practical importance. Our contribution is also two-fold: I) we designed QML models and analyzed their expressive power via their local effective dimension, since their local effective dimension correlates with their classification capability, and II) we proposed and utilized quantum transfer learning for benchmarking and analyzing QML models on real-world, hard-to-classify datasets, because weaker models generate similar performance metrics (e.g., classification accuracy or loss) on real-world, easy-to-classify datasets as more powerful models.

For ongoing and future work, we will integrate faster and simpler QML models with artificial intelligence methodologies based on their quantum resource required \cite{otgonbaatar2023exploiting}. Plus, we will invent and design quantum-inspired networks for practical and significant problems to obtain quantum advantage as early and efficiently as possible. More importantly, quantum and quantum-inspired models help boost conventional probabilistic models for remotely-sensed datasets. 
In addition to the computational advantage of a quantum computer, another advantage of quantum computing to remote sensing is that quantum machine learning algorithms are operate inherently on complex vector space \cite{glasser2019expressive}, and some remote sensing datasets are complex-numbered images. Therefore, we design inherently complex quantum machine learning algorithms for complex-numbered remote sensed images like synthetic aperture radar (SAR) images without the need of modification in conventional machine and deep learning techniques operating on real number space.

\section*{Acknowledgments}
We would like to acknowledge Begüm Demir (Technical University of Berlin) for her valuable comments and suggestions on the very first draft of this paper.

\bibliography{IEEEabrv,qtransfer}

\begin{thebibliography}{10}
\providecommand{\url}[1]{#1}
\csname url@samestyle\endcsname
\providecommand{\newblock}{\relax}
\providecommand{\bibinfo}[2]{#2}
\providecommand{\BIBentrySTDinterwordspacing}{\spaceskip=0pt\relax}
\providecommand{\BIBentryALTinterwordstretchfactor}{4}
\providecommand{\BIBentryALTinterwordspacing}{\spaceskip=\fontdimen2\font plus
\BIBentryALTinterwordstretchfactor\fontdimen3\font minus \fontdimen4\font\relax}
\providecommand{\BIBforeignlanguage}[2]{{%
\expandafter\ifx\csname l@#1\endcsname\relax
\typeout{** WARNING: IEEEtran.bst: No hyphenation pattern has been}%
\typeout{** loaded for the language `#1'. Using the pattern for}%
\typeout{** the default language instead.}%
\else
\language=\csname l@#1\endcsname
\fi
#2}}
\providecommand{\BIBdecl}{\relax}
\BIBdecl

\bibitem{biamonte}
\BIBentryALTinterwordspacing
J.~Biamonte, P.~Wittek, N.~Pancotti, P.~Rebentrost, N.~Wiebe, and S.~Lloyd, ``Quantum machine learning,'' \emph{Nature}, vol. 549, no. 7671, pp. 195--202, Sep 2017. [Online]. Available: \url{https://doi.org/10.1038/nature23474}
\BIBentrySTDinterwordspacing

\bibitem{qpca}
\BIBentryALTinterwordspacing
S.~Lloyd, M.~Mohseni, and P.~Rebentrost, ``Quantum principal component analysis,'' \emph{Nature Physics}, vol.~10, no.~9, pp. 631--633, Sep 2014. [Online]. Available: \url{https://doi.org/10.1038/nphys3029}
\BIBentrySTDinterwordspacing

\bibitem{dunjko}
\BIBentryALTinterwordspacing
V.~Dunjko and H.~J. Briegel, ``Machine learning {\&} artificial intelligence in the quantum domain: a review of recent progress,'' \emph{Reports on Progress in Physics}, vol.~81, no.~7, p. 074001, Jun 2018. [Online]. Available: \url{https://doi.org/10.1088/1361-6633/aab406}
\BIBentrySTDinterwordspacing

\bibitem{qinspired}
\BIBentryALTinterwordspacing
E.~M. Stoudenmire and D.~J. Schwab, ``Supervised learning with quantum-inspired tensor networks,'' 2016. [Online]. Available: \url{https://arxiv.org/abs/1605.05775}
\BIBentrySTDinterwordspacing

\bibitem{Felser_qinspired}
\BIBentryALTinterwordspacing
T.~Felser, M.~Trenti, L.~Sestini, A.~Gianelle, D.~Zuliani, D.~Lucchesi, and S.~Montangero, ``Quantum-inspired machine learning on high-energy physics data,'' \emph{npj Quantum Information}, vol.~7, no.~1, p. 111, Jul 2021. [Online]. Available: \url{https://doi.org/10.1038/s41534-021-00443-w}
\BIBentrySTDinterwordspacing

\bibitem{Zheng2021}
\BIBentryALTinterwordspacing
P.~Zheng, R.~Zubatyuk, W.~Wu, O.~Isayev, and P.~O. Dral, ``Artificial intelligence-enhanced quantum chemical method with broad applicability,'' \emph{Nature Communications}, vol.~12, no.~1, p. 7022, Dec 2021. [Online]. Available: \url{https://doi.org/10.1038/s41467-021-27340-2}
\BIBentrySTDinterwordspacing

\bibitem{WILLSCH2020}
\BIBentryALTinterwordspacing
D.~Willsch, M.~Willsch, H.~{De Raedt}, and K.~Michielsen, ``Support vector machines on the \textit{D-Wave} quantum annealer,'' \emph{Computer Physics Communications}, vol. 248, p. 107006, 2020. [Online]. Available: \url{https://www.sciencedirect.com/science/article/pii/S001046551930342X}
\BIBentrySTDinterwordspacing

\bibitem{Tuysuz2021}
\BIBentryALTinterwordspacing
C.~T{\"u}ys{\"u}z, C.~Rieger, K.~Novotny, B.~Demirk{\"o}z, D.~Dobos, K.~Potamianos, S.~Vallecorsa, J.-R. Vlimant, and R.~Forster, ``Hybrid quantum classical graph neural networks for particle track reconstruction,'' \emph{Quantum Machine Intelligence}, vol.~3, no.~2, p.~29, Nov 2021. [Online]. Available: \url{https://doi.org/10.1007/s42484-021-00055-9}
\BIBentrySTDinterwordspacing

\bibitem{sozogate}
S.~Otgonbaatar and M.~Datcu, ``Classification of remote sensing images with parameterized quantum gates,'' \emph{IEEE Geoscience and Remote Sensing Letters}, vol.~19, pp. 1--5, 2022.

\bibitem{Abbas2021}
\BIBentryALTinterwordspacing
A.~Abbas, D.~Sutter, C.~Zoufal, A.~Lucchi, A.~Figalli, and S.~Woerner, ``The power of quantum neural networks,'' \emph{Nature Computational Science}, vol.~1, no.~6, pp. 403--409, Jun 2021. [Online]. Available: \url{https://doi.org/10.1038/s43588-021-00084-1}
\BIBentrySTDinterwordspacing

\bibitem{abbaslocal}
\BIBentryALTinterwordspacing
A.~Abbas, D.~Sutter, A.~Figalli, and S.~Woerner, ``Effective dimension of machine learning models,'' 2021. [Online]. Available: \url{https://arxiv.org/abs/2112.04807}
\BIBentrySTDinterwordspacing

\bibitem{qsvm}
\BIBentryALTinterwordspacing
P.~Rebentrost, M.~Mohseni, and S.~Lloyd, ``Quantum support vector machine for big data classification,'' \emph{Phys. Rev. Lett.}, vol. 113, p. 130503, Sep 2014. [Online]. Available: \url{https://link.aps.org/doi/10.1103/PhysRevLett.113.130503}
\BIBentrySTDinterwordspacing

\bibitem{preskill}
\BIBentryALTinterwordspacing
J.~Preskill, ``Quantum {C}omputing in the {NISQ} era and beyond,'' \emph{{Quantum}}, vol.~2, p.~79, Aug. 2018. [Online]. Available: \url{https://doi.org/10.22331/q-2018-08-06-79}
\BIBentrySTDinterwordspacing

\bibitem{schuld}
\BIBentryALTinterwordspacing
M.~Schuld and N.~Killoran, ``Quantum machine learning in feature hilbert spaces,'' \emph{Phys. Rev. Lett.}, vol. 122, p. 040504, Feb 2019. [Online]. Available: \url{https://link.aps.org/doi/10.1103/PhysRevLett.122.040504}
\BIBentrySTDinterwordspacing

\bibitem{eurosat}
P.~Helber, B.~Bischke, A.~Dengel, and D.~Borth, ``Eurosat: A novel dataset and deep learning benchmark for land use and land cover classification,'' \emph{IEEE Journal of Selected Topics in Applied Earth Observations and Remote Sensing}, vol.~12, no.~7, pp. 2217--2226, 2019.

\bibitem{copy1}
A.~Sebastianelli, D.~A. Zaidenberg, D.~Spiller, B.~L. Saux, and S.~L. Ullo, ``On circuit-based hybrid quantum neural networks for remote sensing imagery classification,'' \emph{IEEE Journal of Selected Topics in Applied Earth Observations and Remote Sensing}, vol.~15, pp. 565--580, 2022.

\bibitem{sozotgrs}
S.~Otgonbaatar and M.~Datcu, ``Natural embedding of the \emph{Stokes} parameters of polarimetric synthetic aperture radar images in a gate-based quantum computer,'' \emph{IEEE Transactions on Geoscience and Remote Sensing}, vol.~60, pp. 1--8, 2022.

\bibitem{schuld_qtf}
\BIBentryALTinterwordspacing
A.~Mari, T.~R. Bromley, J.~Izaac, M.~Schuld, and N.~Killoran, ``Transfer learning in hybrid classical-quantum neural networks,'' \emph{{Quantum}}, vol.~4, p. 340, Oct. 2020. [Online]. Available: \url{https://doi.org/10.22331/q-2020-10-09-340}
\BIBentrySTDinterwordspacing

\bibitem{datareuploading}
\BIBentryALTinterwordspacing
A.~P{\'{e}}rez-Salinas, A.~Cervera-Lierta, E.~Gil-Fuster, and J.~I. Latorre, ``Data re-uploading for a universal quantum classifier,'' \emph{{Quantum}}, vol.~4, p. 226, Feb. 2020. [Online]. Available: \url{https://doi.org/10.22331/q-2020-02-06-226}
\BIBentrySTDinterwordspacing

\bibitem{seth}
\BIBentryALTinterwordspacing
S.~Lloyd, M.~Schuld, A.~Ijaz, J.~Izaac, and N.~Killoran, ``Quantum embeddings for machine learning,'' May 2020. [Online]. Available: \url{https://arxiv.org/abs/2001.03622}
\BIBentrySTDinterwordspacing

\bibitem{vgg16}
\BIBentryALTinterwordspacing
K.~Simonyan and A.~Zisserman, ``Very deep convolutional networks for large-scale image recognition,'' in \emph{3rd International Conference on Learning Representations, {ICLR} 2015, San Diego, CA, USA, May 7-9, 2015, Conference Track Proceedings}, Y.~Bengio and Y.~LeCun, Eds., 2015. [Online]. Available: \url{http://arxiv.org/abs/1409.1556}
\BIBentrySTDinterwordspacing

\bibitem{schuldnlayer}
\BIBentryALTinterwordspacing
M.~Schuld, A.~Bocharov, K.~M. Svore, and N.~Wiebe, ``Circuit-centric quantum classifiers,'' \emph{Phys. Rev. A}, vol. 101, p. 032308, Mar 2020. [Online]. Available: \url{https://link.aps.org/doi/10.1103/PhysRevA.101.032308}
\BIBentrySTDinterwordspacing

\bibitem{Haug_2021}
\BIBentryALTinterwordspacing
T.~Haug, K.~Bharti, and M.~Kim, ``Capacity and quantum geometry of parametrized quantum circuits,'' \emph{PRX Quantum}, vol.~2, p. 040309, Oct 2021. [Online]. Available: \url{https://link.aps.org/doi/10.1103/PRXQuantum.2.040309}
\BIBentrySTDinterwordspacing

\bibitem{classicallocal}
\BIBentryALTinterwordspacing
W.~J. Maddox, G.~Benton, and A.~G. Wilson, ``Rethinking parameter counting in deep models: Effective dimensionality revisited,'' 2020. [Online]. Available: \url{https://arxiv.org/abs/2003.02139}
\BIBentrySTDinterwordspacing

\bibitem{yuxuan_2020}
\BIBentryALTinterwordspacing
Y.~Du, M.-H. Hsieh, T.~Liu, and D.~Tao, ``Expressive power of parametrized quantum circuits,'' \emph{Phys. Rev. Research}, vol.~2, p. 033125, Jul 2020. [Online]. Available: \url{https://link.aps.org/doi/10.1103/PhysRevResearch.2.033125}
\BIBentrySTDinterwordspacing

\bibitem{cheng0}
G.~Cheng, X.~Xie, J.~Han, L.~Guo, and G.-S. Xia, ``Remote sensing image scene classification meets deep learning: Challenges, methods, benchmarks, and opportunities,'' \emph{IEEE Journal of Selected Topics in Applied Earth Observations and Remote Sensing}, vol.~13, pp. 3735--3756, 2020.

\bibitem{cheng1}
G.~Cheng, J.~Han, and X.~Lu, ``Remote sensing image scene classification: Benchmark and state of the art,'' \emph{Proceedings of the IEEE}, vol. 105, no.~10, pp. 1865--1883, 2017.

\bibitem{pennylane}
\BIBentryALTinterwordspacing
V.~Bergholm, J.~Izaac, M.~Schuld, C.~Gogolin, S.~Ahmed, V.~Ajith, M.~S. Alam, G.~Alonso-Linaje, B.~AkashNarayanan, A.~Asadi, J.~M. Arrazola, U.~Azad, S.~Banning, C.~Blank, T.~R. Bromley, B.~A. Cordier, J.~Ceroni, A.~Delgado, O.~Di~Matteo, A.~Dusko, T.~Garg, D.~Guala, A.~Hayes, R.~Hill, A.~Ijaz, T.~Isacsson, D.~Ittah, S.~Jahangiri, P.~Jain, E.~Jiang, A.~Khandelwal, K.~Kottmann, R.~A. Lang, C.~Lee, T.~Loke, A.~Lowe, K.~McKiernan, J.~J. Meyer, J.~A. Montañez-Barrera, R.~Moyard, Z.~Niu, L.~J. O'Riordan, S.~Oud, A.~Panigrahi, C.-Y. Park, D.~Polatajko, N.~Quesada, C.~Roberts, N.~Sá, I.~Schoch, B.~Shi, S.~Shu, S.~Sim, A.~Singh, I.~Strandberg, J.~Soni, A.~Száva, S.~Thabet, R.~A. Vargas-Hernández, T.~Vincent, N.~Vitucci, M.~Weber, D.~Wierichs, R.~Wiersema, M.~Willmann, V.~Wong, S.~Zhang, and N.~Killoran, ``\textit{PennyLane}: Automatic differentiation of hybrid quantum-classical computations,'' 2018. [Online]. Available: \url{https://arxiv.org/abs/1811.04968}
\BIBentrySTDinterwordspacing

\bibitem{Meyer_2021}
\BIBentryALTinterwordspacing
J.~J. Meyer, ``Fisher information in noisy intermediate-scale quantum applications,'' \emph{Quantum}, vol.~5, p. 539, Sep 2021. [Online]. Available: \url{https://doi.org/10.22331%2Fq-2021-09-09-539}
\BIBentrySTDinterwordspacing

\bibitem{FIM}
\BIBentryALTinterwordspacing
A.~Ly, M.~Marsman, J.~Verhagen, R.~Grasman, and E.-J. Wagenmakers, ``A tutorial on \textit{Fisher} information,'' 2017. [Online]. Available: \url{https://arxiv.org/abs/1705.01064}
\BIBentrySTDinterwordspacing

\bibitem{sozomdpi}
\BIBentryALTinterwordspacing
S.~Otgonbaatar and M.~Datcu, ``Assembly of a coreset of earth observation images on a small quantum computer,'' \emph{Electronics}, vol.~10, no.~20, 2021. [Online]. Available: \url{https://www.mdpi.com/2079-9292/10/20/2482}
\BIBentrySTDinterwordspacing

\bibitem{Huang2021}
\BIBentryALTinterwordspacing
H.-Y. Huang, M.~Broughton, M.~Mohseni, R.~Babbush, S.~Boixo, H.~Neven, and J.~R. McClean, ``Power of data in quantum machine learning,'' \emph{Nature Communications}, vol.~12, no.~1, p. 2631, May 2021. [Online]. Available: \url{https://doi.org/10.1038/s41467-021-22539-9}
\BIBentrySTDinterwordspacing

\bibitem{McClean2018}
\BIBentryALTinterwordspacing
J.~R. McClean, S.~Boixo, V.~N. Smelyanskiy, R.~Babbush, and H.~Neven, ``Barren plateaus in quantum neural network training landscapes,'' \emph{Nature Communications}, vol.~9, no.~1, p. 4812, Nov 2018. [Online]. Available: \url{https://doi.org/10.1038/s41467-018-07090-4}
\BIBentrySTDinterwordspacing

\bibitem{otgonbaatar2023exploiting}
S.~Otgonbaatar and D.~Kranzlmüller, ``Exploiting the quantum advantage for satellite image processing: Quantum resource estimation,'' 2023.

\bibitem{glasser2019expressive}
I.~Glasser, R.~Sweke, N.~Pancotti, J.~Eisert, and J.~I. Cirac, ``Expressive power of tensor-network factorizations for probabilistic modeling, with applications from hidden markov models to quantum machine learning,'' 2019.

\end{thebibliography}
\bibliographystyle{IEEEtran}

\end{document}